# Multiuser Cellular Network

Yi Bao, Chao Wang, Ming Chen
*University of Science and Technology of China*

## Abstract

Modern radio communication is faced with a problem about how to distribute restricted frequency to users in a certain space. Since our task is to minimize the number of repeaters, a natural idea is enlarging coverage area. However, coverage has restrictions. First, service area has to be divided economically as repeater's coverage is limited. In this paper, our fundamental method is to adopt seamless cellular network division. Second, underlying physics content in frequency distribution problem is interference between two close frequencies. Consequently, we choose a proper frequency width of 0.1MHz and a relevantly reliable setting to apply one frequency several times.

We make a few general assumptions to simplify real situation. For instance, immobile users yield to homogenous distribution; repeaters can receive and transmit information in any given frequency in duplex operation; coverage is mainly decided by antenna height.

Two models are built up to solve 1000 users and 10000 users situations respectively. In order to utilize restricted frequency and PL code, three stratified terms - "cell", "cluster", "group" – are introduced to describe the models in detail. Under our analysis, 91 repeaters for 1000 users and 469 repeaters for 10000 users are viable results.

Next, to test stability and sensitivity of models, we give total consideration to the variation of sum of users, antenna height, and frequency width and service radius. Evaluation about models is offered qualitatively. Finally, two practical cases are put forward to gain a partial knowledge of mountainous area. The brief method in dealing with mountains is classified discussion in two ideal conditions. It may provide some constructive suggestions to avoid shortcomings or take proper measures in similar locations.

# Table of Contents

# Introduction

Communication is important throughout the history. From the beacon towers to the mobile phone, people keep finding ways to exchange messages with each other faster and faster. With the application of radio, communication has been developing in tremendous speed. VHF is a radio spectrum range from 30MHz to 300MHz, frequently used in FM radio broadcast, television broadcast, land mobile stations (emergency, business, and military), long range data communication with radio modems, Amateur Radio, marine communications, air traffic control communications and air navigation systems (e.g. VOR, DME & ILS). Together with the enhancement of communication speed, distance that signal can cover also get improved with the appearance of repeaters.

However, interference between adjacent repeaters which transmits approaching frequencies cannot be overlooked. In order to mitigate the channel interference problem, CTCSS, also called PL, is brought in, superimposing a given sub-audible tone on the transmitted signal, giving a characteristic frequency to each request call without interference.

Given the problem to arrange fewest repeaters among the area that can afford all the users' communication demand, we are faced with the following questions:
- **How to divide the area**: the coverage of repeater is limited.
- **What is the shortest bandwidth between adjacent frequencies without confusion**: Monkey chatter should be avoided.
- **The usage of 54 different PL tones**: the way in which PL tones is divided determines the validity of transmission.
- **How to distinguish every user**
- **How do repeaters function to transmit information**

The assumptions of our model are as below:

# Assumptions

Based on analysis of the given conditions, we find that variations of repeaters' quality, users, identity categories, etc have to be simplified to accommodate a virtual situation, such as the first question about 1000 users. As a result, several basic assumptions must be made in this step as below. It is quite important to keep in mind that these assumptions may prove to be premature and some revises will probably be added into the model in the subsequent steps.
- **The terrain of the area is a flat landscape without mountains or other projections.** For

most of the time, space we are discussing can be regarded in a quasi-two-dimension area.

- **1000 users yield to homogeneous distribution.** Every user may appear in any position in the circuit for equal probability.
- **Every user has a specific "ID".** An ID includes one frequency, which is called a characteristic frequency, and one sub-audible tone.
- **Users are forbidden to hold carrier frequencies that are too close or stay too close.** In other words, we assume that a variable Δf to divide the frequency range and suitable space interval are needed.
- **One repeater is described as one antenna and each repeater has the same coverage.** We also notice that in plentiful cases, several repeaters share a separate antenna. However, the site area in the first question has already been regarded as flat and the number of users is far from crowded. A separate antenna includes a transmitter and a receiver, its tentative radius is 5 miles.
- **Antennae have same input and output quality among all directions.**
- **Given area should be divided seamlessly.** Cellular beehive division, also called hexagon division, turns out to be suitable in this case.

To achieve the goal that every position among the given area should be covered by at least one repeater, a lot of circles are needed. It is easily imagined that circles will overlap with each other to some extent, giving consideration to the fact; valid areas are equilateral polygons as below. Mathematics demonstrates that only three shapes satisfying the condition: regular triangle, square and regular hexagon.

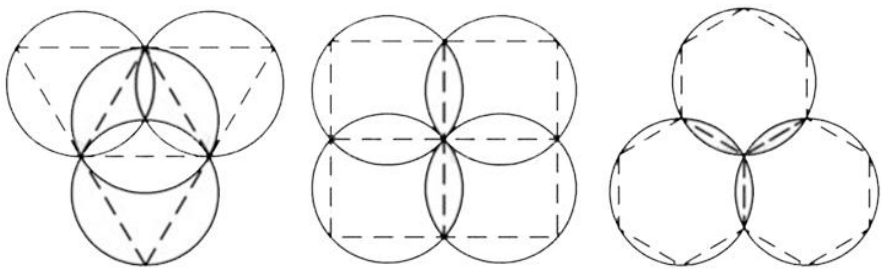

- Assuming that every circle has equal radius r, which kind of polygon division has the highest valid area to make use of limited coverage? Chart below shows that hexagon division takes most advantage of repeater coverage among the three.

| Shape | Regular triangle | Square | Regular hexagon |
|---|---|---|---|
| Distance between two shapes | R | r | $\sqrt{3}\,r$ |
| Valid area | $1.3\,r^2$ | $2\,r^2$ | $2.6\,r^2$ |
| Overlap area | $1.2\pi r^2$ | $0.73\pi r^2$ | $0.35\pi r^2$ |

- **Each hexagon has exactly one repeater, which amplifies the information from the hexagon or work as a way station to transmit signal, in the right center.** Coverage is mainly decided by the height of antenna. Although other factors, such as output power, gain, obliquity, will affect the coverage in real situation, we suppose that it would be very complex to take all of them into consideration at the same time. Furthermore, they will not change the modeling process too much as an essential problem is about coverage area only.
- **Every repeater is in full duplex operation and more than one user can get an access to the repeater simultaneously.** In communication of a user-to-user line, every individual

- repeater involved is able to proceed transmitting and receiving at the same time.
- **Users are stationary in geography.**
- **All users are able to transmit all the 42 PL code in message information.** It does not go against with that each has one characteristic PL.

# The Model

## Definitions

- R---the radius of the area supplied with radio service
- r ---the radius of a repeater's coverage for signals coming from users(small signals)
- R'---the radius of a repeater's coverage for signals coming from repeaters (amplified signals)
- N---the number of clusters
- $n_c$---the number of cells in one cluster
- n---a cluster's capacity to offer radio service
- Δf---the minimum frequency bandwidth between neighboring frequencies without interference

## Cellular Network

Considering the coverage overlap by each repeaters, the area is divided into regular hexagonal cells .Several cells make up a cluster (the number of cells $n_c$ satisfy the identity $n_c=i^2+ij+j^2$ ∀ i ,j ∈N   i+j≠0).Each   cluster   is   assigned   a   specific   PL   tone  .The   series   of frequencies(145MHz-147.4MHz)[1] are available to all the clusters without interference and being shared alike among the cells in one cluster . The radius of the coverage of a repeater is under the limit of the height of antenna and the power of signal .For the signals without amplified ,they can only receive the repeater at the center of the hexagonal cell , while amplified signals can reach to its neighboring repeaters . Obtained by geometry $R'=\sqrt{3}r$, $n_L$ is deduced by the recursion formula

$$r\sqrt{3(n_L-1)^2+1} \leq R \leq r\sqrt{3n_L^2+1}$$

.

---

[1] the frequency bandwidth is set to satisfy condition in problem "the transmitter frequency in a repeater is either 600 kHz above or 600 kHz below the receiver frequency

The capacity of each cluster is under the constrain of $\Delta f$, $n = 2.4MHz/\Delta f$, $N = 1000/n$.

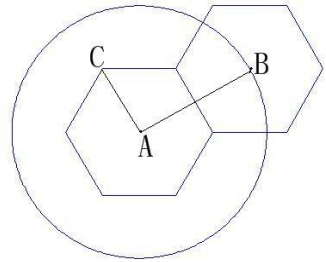

$r=||AC||$ $R'=||AB||$

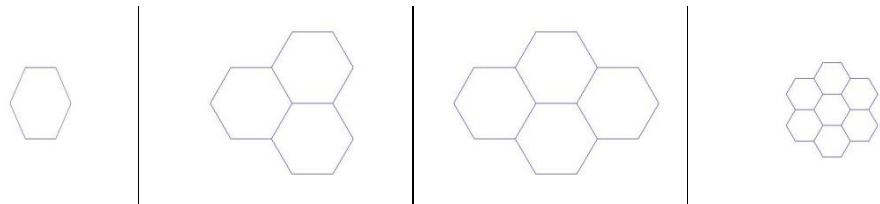

Several cells make up a cluster. The number of cells $n_c$ in one cluster satisfies identity $n_c=i^2+ij+j^2$

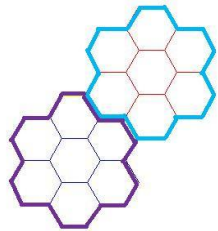

The series of frequencies are available to all clusters without interference

## Communication

Communications can be set up in the different cells or clusters. Specific frequency and PL tone make users distinguished. When X wants to communicate with Y, he first sends a Y's characteristic frequency (or Y's characteristic frequency plus 600Hz, to be explained later) conveys information and Y's PL tone, superimposing its own PL tone, to its center repeater. Suppose every repeater has an adequate storage to memorize the next step towards destination's repeater. Repeaters only receive and process signal carrying their specific PL tone. When dealing with the same call request from different users or repeaters at the same time, repeater sets up a stack, which guarantees first request first handled while the latter ones have to wait.

*Let us make an interesting comparison to demonstrate how repeaters work. To set up a communication channel between two users is akin to the way of post system works. Route has been set. We write down digital information in the letter, put it in a special envelop, the color of envelop represents the target's PL tone, you can get recipient's address (target user's characteristic frequency) and return*

*address(request user's characteristic frequency) from the envelop .Every transfer post office sticks a stamp(transfer repeater's PL tone) on the envelop which can only afford to get to the next post office along the route till the target post office. Post office only distributes letters with certain color of envelopes which represents the region in its charge , otherwise sticks new stamps and passes them on.*

*You can also understand it in this way . All of the repeaters' PL tones are recorded on a long enough belt with many gears ,gears on part of the belt stand for a repeater's PL tone .When some road is requested to open ,a certain wheel with gears recorded with route is used. Only in the fit way can the wheel advances .The wheel can bring essential information(transfer repeaters' PL tone and target's characteristic frequency) about the route together with some other information to mark the source of the request(request user's characteristic frequency ). With the wheel going forwards, signals transmit.*

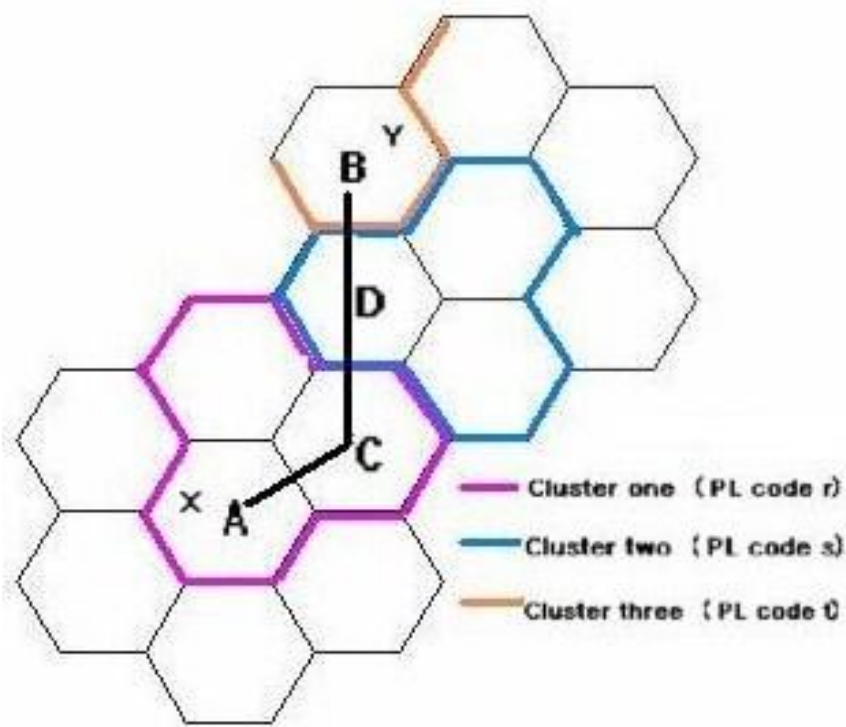

Suppose repeater only processes signals with the cluster's specific PL tone but can retransmit any neighboring PL tones, any possible used frequencies in its effective coverage can receive and transmit. We set X, Y to represent users and users' specific frequencies without confusion .

For the communication between X and Y ,a suitable route has already been designed and memorized, given that X->A->C->D->B->Y, communication channel builds up in this way:

> User X sends a signal to repeater A ,with the transmitter frequency Y&Y PL, superimposing PL tone r (The first step's transmitter frequency is Y or Y+600Hz is determined by whether the total steps is odd or even)

> A receives the signal and processes X 's request ,amplifying the signal and retransmitting to repeater C, with the transmitter frequency Y+600Hz

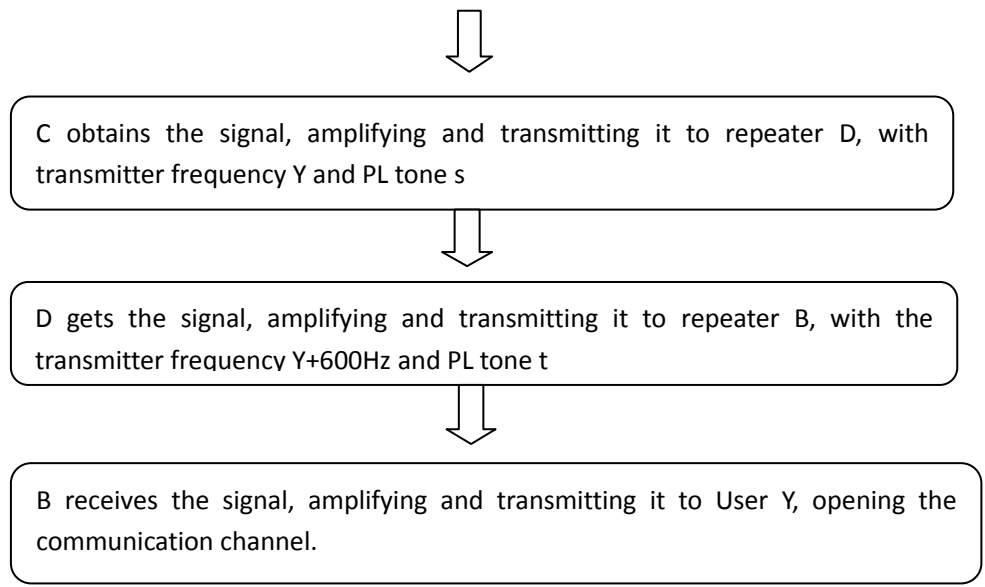

C obtains the signal, amplifying and transmitting it to repeater D, with transmitter frequency Y and PL tone s

D gets the signal, amplifying and transmitting it to repeater B, with the transmitter frequency Y+600Hz and PL tone t

B receives the signal, amplifying and transmitting it to User Y, opening the communication channel.

# Tentative Calculation

## Determination about repeater coverage

No matter how powerful a repeater is, its maximum radius is limited by line-of-sight. The critical situation of is given below:

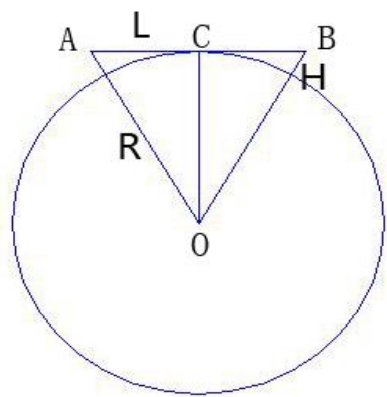

- A and B--- the ends of two antennae
- O---the earth's core
- C---line-of-sight farthest point
- H---antenna height
- R---radius of the earth
- L---maximum coverage radius

$$L = \sqrt{(R+H)^2 - R^2} \approx \sqrt{2RH} = \sqrt{2 \times 6378 \times 10^3 H} \approx 3571.55\sqrt{H}$$

Another way to estimate the maximum radius of an antenna results from experience:

$$r = \sqrt{1.5H}$$

Here R is in miles and H is in feet.

In assumptions, we regard antenna height as the main factor to affect repeater coverage. In terms of that it is a flat area and the entire area is not so big, we set antenna height as 15 meters (about 49.2 feet, which means an antenna, may be set up on the top of a tower). We take the two equations above into calculation and get nearly the same answer, 8.7 miles, which has been used in the following steps.

## Principle of area separation

To cover the area of radius in 40 miles, hexagons should be assigned evenly on the map. For example, the total number of hexagons should be 1, 7, 19, 37…

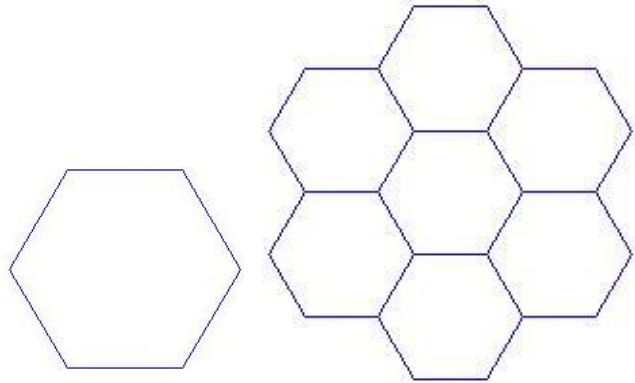

The graph of relation between number of cells and coverage radius of one repeater is shown as below:

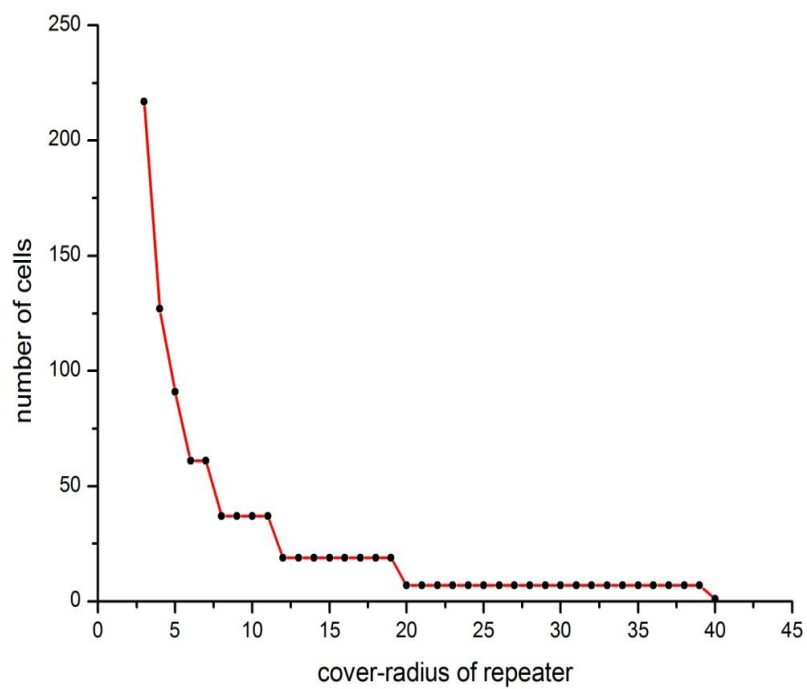

In this case, the result turns out to be 91 hexagons(91 repeaters):

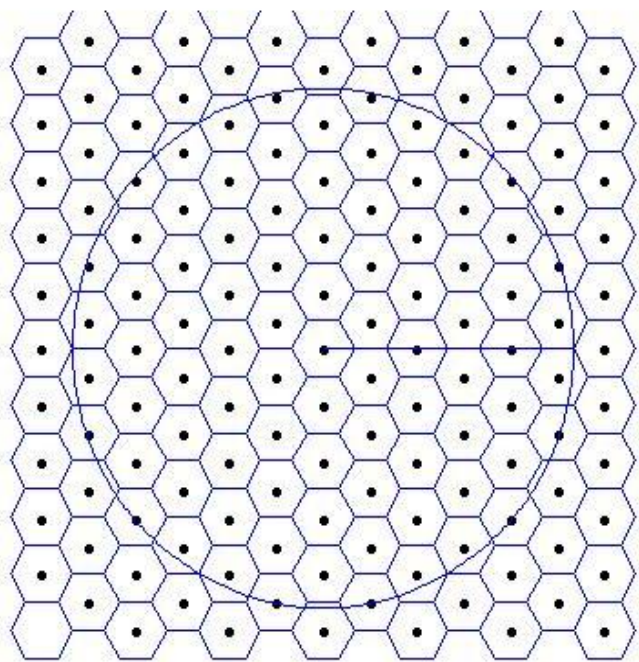

Black points above in the center of each hexagon represent repeaters. It is apparent that some area of the 91 hexagons are out of the circle.

As we discuss above, in terms of two factors: frequency width between 145 to 148 MHz and frequency deviation of 600 kHz, we assume that valid frequency width which can be used for carrier frequency is between 145 and 147.4 MHz. Then, Δf is assumed to be 0.1MHz (an empirical minimum frequency width). Consequently we obtain:

$$n = \frac{2.4 MHz}{\Delta f} = 24$$

Now every cluster can gain 24 frequencies, which not only are different in frequency numbers but also will absolutely not interfere with each other, to distribute into its cells. It means that one cluster has a capacity of 24 different users.

For the first question, 1000 users are distributed in the circle. The number of clusters is decided as:

$$N \geq \left[\frac{1000}{n} + 1\right]$$

Hence, 1000 users will be distributed into 42 clusters (42 PL tones have been used) at least.

## A feasible solution

Naively, we set that one cluster includes three cells or one cell. We set x as the number of clusters including three cells and y as the number of clusters including one cell. Then an inequality system is given as below:

$$\begin{cases} 3x + y \geq 91 \\ x + y = 42 \end{cases}$$

A reasonable integer solution is x=25, y=17. There does not exist an exclusive method for actual mapping and one of them has been shown below:

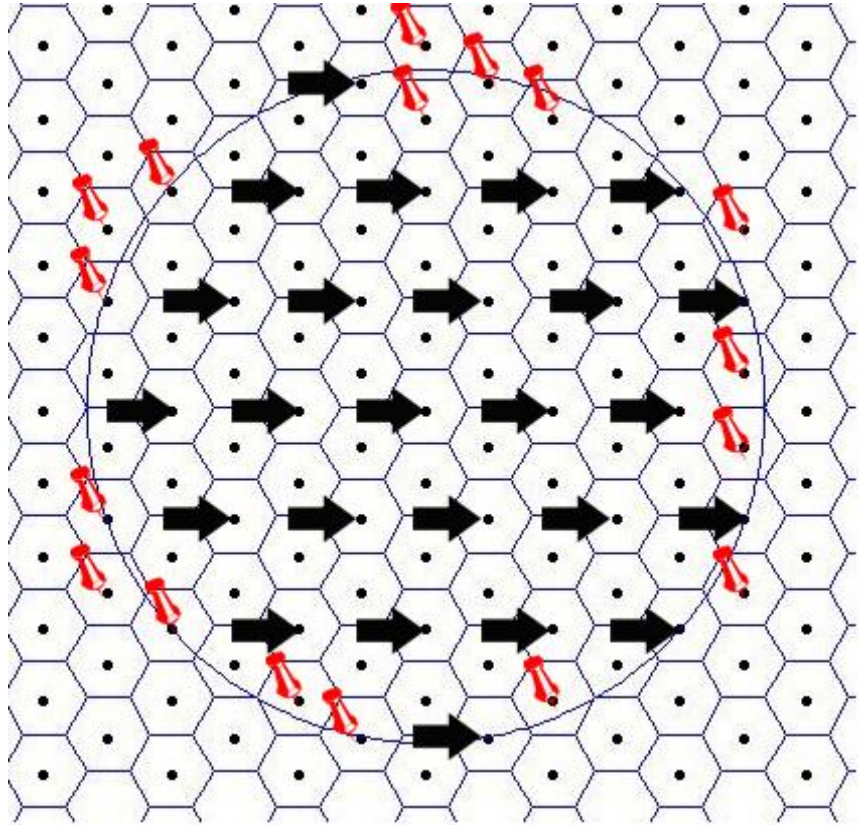

Black arrows represent the center of clusters which have three cells while red pins represent clusters having one cell. Careful readers may find that the topmost black arrow consists of one needless cell; therefore the valid number of repeaters is still 91.

# 10,000-user-model

When the number of users turns into 10000, the model must change. In this model, the capacity of each cluster is under the constrain of $\Delta f$, $n = \dfrac{2.4MHz}{\Delta f}$, $N = 10000/n$

We assume: $\Delta f = 0.1MHz$

So that $n = \dfrac{2.4MHz}{\Delta f} = \dfrac{2.4MHz}{0.1MHz} = 24$.

And the number of clusters is $N = 10000/n = \dfrac{10000}{24} = 416.7$

On one hand, we need to cover this area seamlessly, so the smallest number of the clusters is 417.

On the other hand, we still have 54 PL tones available. We can combine 54 clusters into one group. In this group, every cluster uses unique PL tone.

The number of these groups we need to cover is defined as Ng.

$$N_g = \frac{416.7}{54} = 7.71.$$

61 hexagonal clusters could make up a regular shape as below:
(One small hexagon means a cluster)

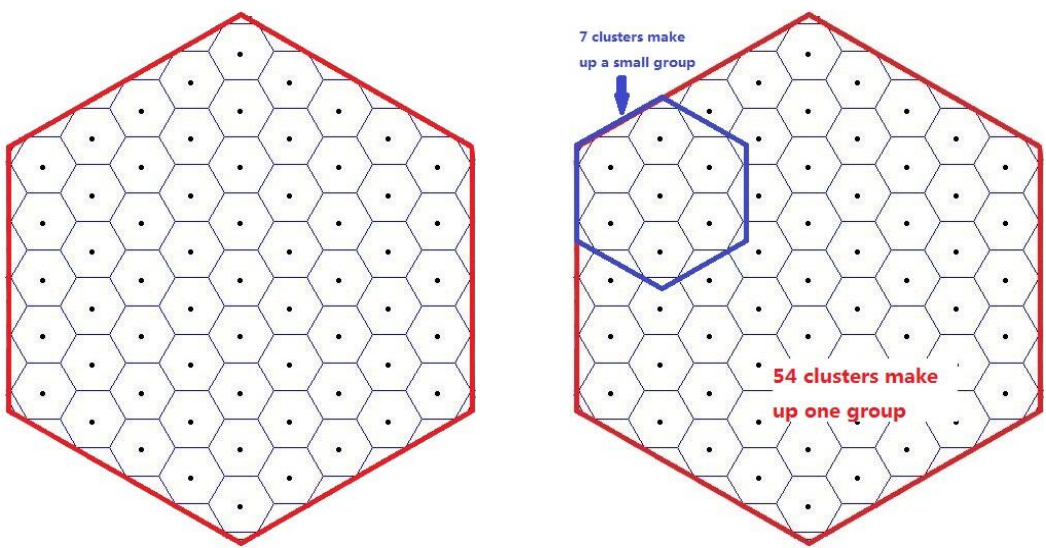

We can choose 54 clusters in this area, leaving 7 clusters remained. The chosen 54 clusters make up one group, and the remaining 7 clusters form another small group. All the 54 PL tones are distributed equally among the chosen 54 clusters. One cluster has one unique PL tone. The small group containing 7 clusters has 7 different PL tones which are picked up from the 54 PL tones. Thus, clusters in one group can be distinguished by different PL tones.

To cover an area of radius 40 miles, 417 clusters are needed at least. Given that the coverage of a repeater is 2 miles, no less than 469 clusters are required. Each of these clusters is under the control of its repeater which is located in the center of the hexagonal area.

For the center part of these 469 clusters, 7 big groups can be combined together, which costs $61 \times 7 = 427$ clusters, while we still need $469 - 427 = 42$ clusters more.

The situation is showed in the following figure:

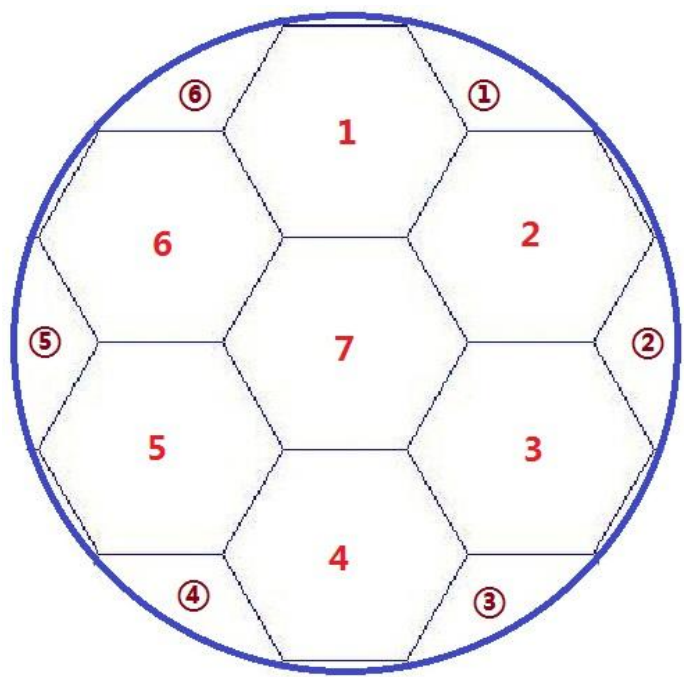

In this figure, groups 1-7 mean the seven big groups that contain 61 clusters. Each one is made up with a 54-cluster-group and a 7-cluster-group.

Groups ① to ⑥ mean six small groups, and each small group contains 7 clusters.

$$7 \times 61 + 6 \times 7 = 4,$$

Namely, the problem to cover area seamlessly has been solved.

Next, we are faced with the question: How to assign unique IDs to the total 10,000 users? There are 24 separated frequency channels between 145 to 147.4 MHz, so each cluster contains 24 users. Every big group signed 1 to 7 contains 1464 users. Every small group at margin contains 168 users. Every big group includes the two different sizes of groups has a unique group ID which we call it group code (GC).

Till now, we know:

There are 20 different GCs in this figure totally:

Every big group (No.1 to No.7) has two different GCs;

Every small group (No. ① to No. ⑥) has one GC;

The total number of GCs is $7 \times 2 + 6 = 20$

With the help of group code, we can distinguish different groups. It means that we can tell any two clusters from each other even if all of their channels and PL tones are the same.

To avoid clusters in different groups who share the same PL tones interfere with each other is of great importance, which puts an requirement on the distribution of PL tones that the two clusters share the same PL tone must be far enough apart.

In our model every 54-cluster-group has the same PL tones distribution.

When it comes to distribute PL tones of the 7-cluster-group in the big group, we make it have a same PL tones distribution as the center part of the 54-cluster-group.

Readers can understand it by the following figure:

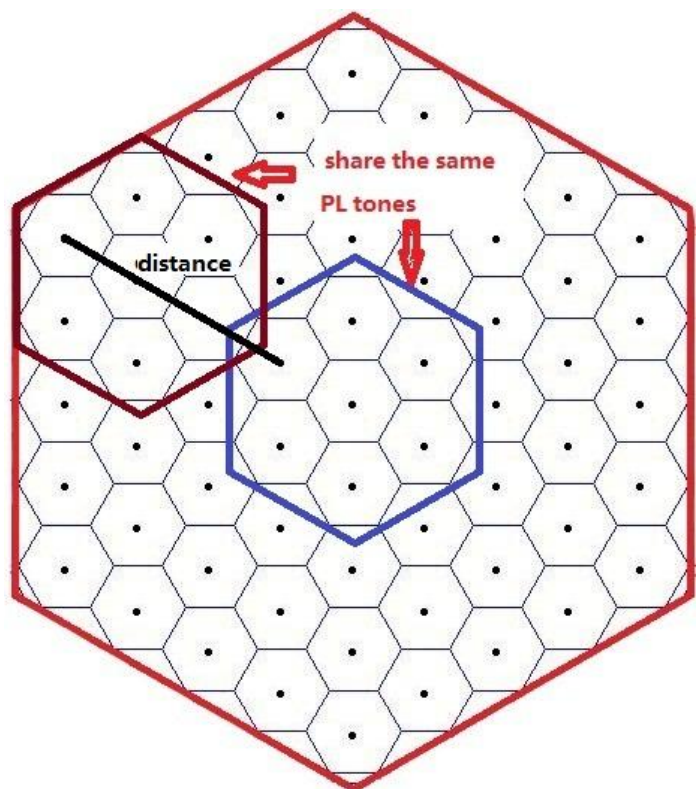

This measure makes all the distance between clusters which have same PL tones being far enough. If the coverage radius of the repeater is 2 miles, this distance is about 10.4 miles.[2]

How to make the division is shown below:

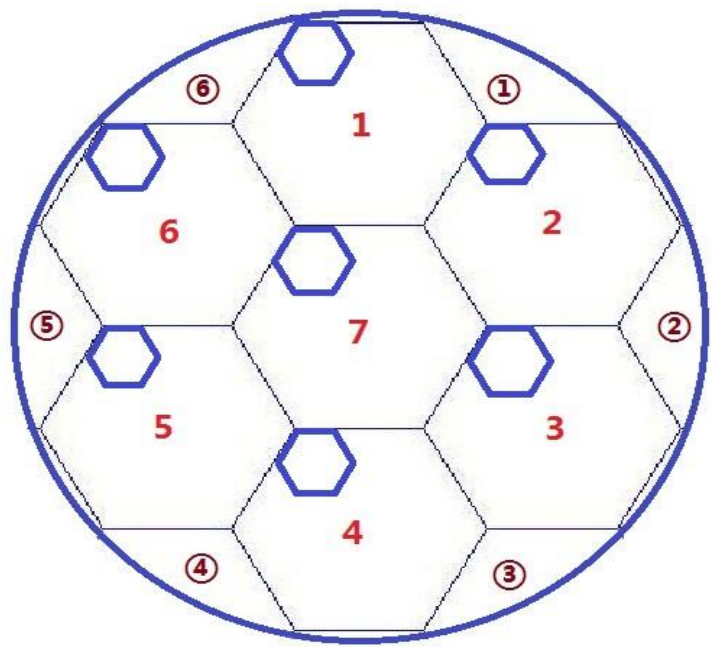

And the PL tone distribution of group No. ① to No. ⑥ is:

---

[2] We assume that if the distance between two clusters which share the same PL tones is more than 10 miles, they would not disturb each other.

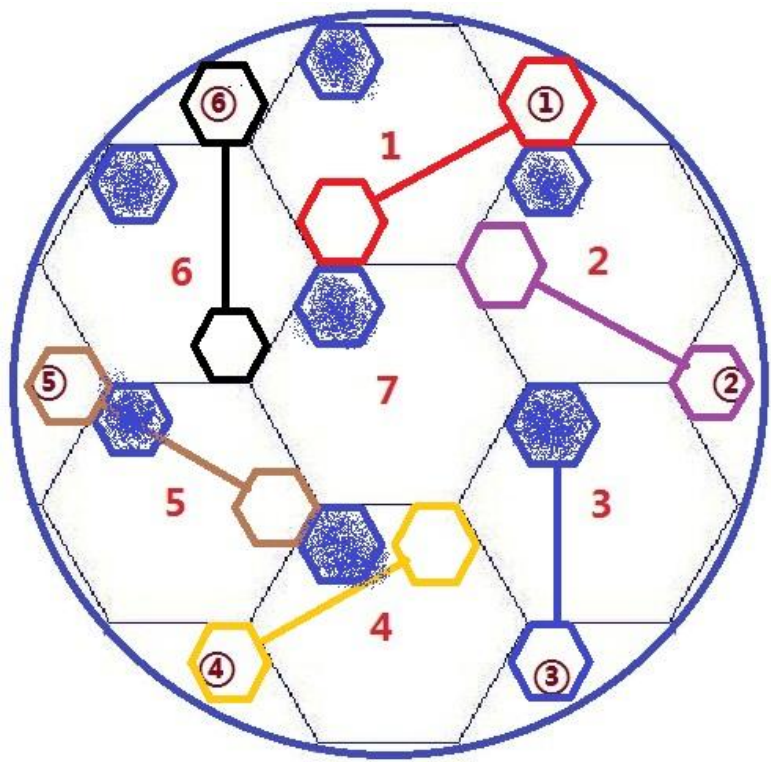

The PL tone distribution among clusters No. ① to No. ⑥ are as same as the areas which are lined with them in color respectively. The distance between them is long enough to mitigate interferences.

The conception of cell does not exist in this model. In addition, the new concept, group, makes the communication between users a little different from the former 1000-user-model. People's ID is unique which is determined by both the PL tone and group code (GC). When X wants to communicate with Y, he must first send the signal with the information of Y's group code and Y's PL tone which is conveyed by the characteristic frequency of Y. The PL tone which is used to transmit between the clusters is superimposed with the special frequency of Y. We still suppose that repeater only processes signals with the cluster's specific PL tone but can retransmit any neighboring PL tones. Every repeater has information about the suitable route between X and Y. But in this case, in order to avoid probability of fault acceptance of cluster which share the same PL tone with original destination cluster, the route cannot include such clusters when designed. For example, if the GC of Y is Q1, and the PL tone is A1, the information mustn't be transmitted through any other cluster which shares the same PL tone A1 as Y in other group. Because the distance between the same-PL-tone cluster is larger than 10 miles, the radio transmits towards the destination will not disturb any others with the same PL tone.

# Mountainous Cases

Mountains or other complicated landscapes often appear in a real design of repeaters coverage.

The uppermost role a mountain plays is to block information ways in the space. Here we still follow the assumption that antenna height is 15 meters.

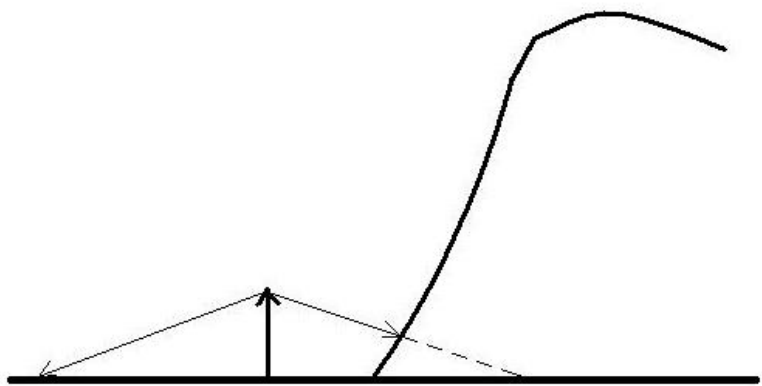

As graph above shows, it is clear that almost all mountains are higher than antenna height remarkably so that the information, whose direction is towards the mountain, will be blocked totally. One possible revise is to add more repeaters at proper positions to maintain necessary connection between original repeaters.

As mountainous situations are too variable to be induced only into one simple model, we discuss mountainous in two cases according to the modes of users. In each case, mountains are divided into two kinds: a large block and a series of small mountains. If mountains in the circle are quite high in altitude (you can call them high flat stands), it is not hard to suppose that several consequent repeaters will be set on the slope path to relay information. We will fix altitude as a constant and ignore repeaters used on the slope in the following analysis, although they will surely increase sum and influence frequency distribution to some extent.

# Case 1: Emergency

If users may appear in the mountains (for example they are a group of field explorers), then repeaters among mountain areas are needed in case of they getting involved into outdoor danger.

## Case 1.A: Small Mountains Straggling

Since users are in an emergency, the repeaters coverage should include all the mountain area to provide effective accesses for them.

It is no harm to assume that there is a small mountain among hexagons as the figure shows below:

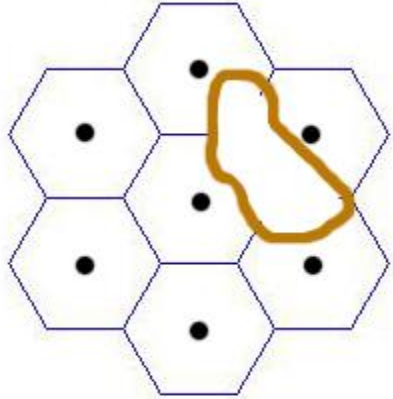

It is obvious that any position of the roof on the small mountain would be available for an antenna site since R' is enough for it to reach at least two repeaters. In other words, just one antenna can satisfy the requirement. Furthermore, this repeater does not need any more restrictions.

## Case 1.B: Large Mountain

Large block of mountain may cover several original repeater positions without difficulty, as graph shows below:

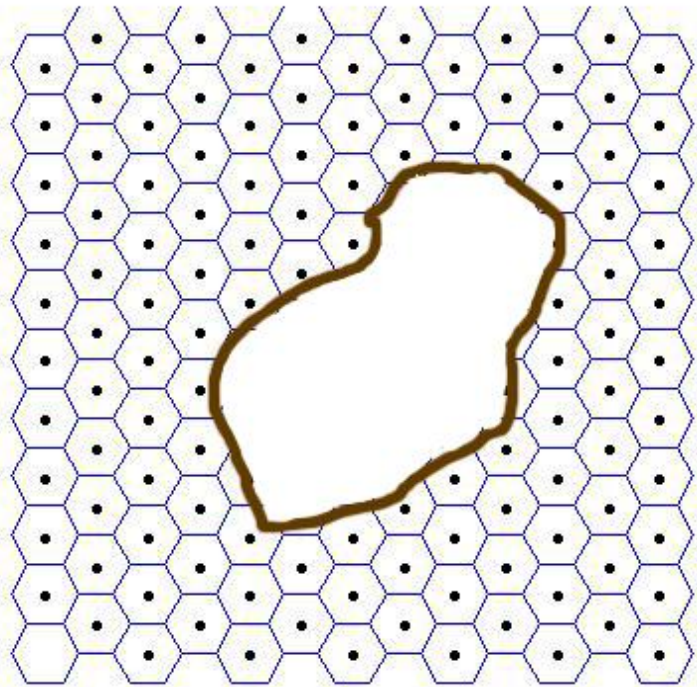

In order to cover overall area on the mountain, a inner hexagon division must be applied:

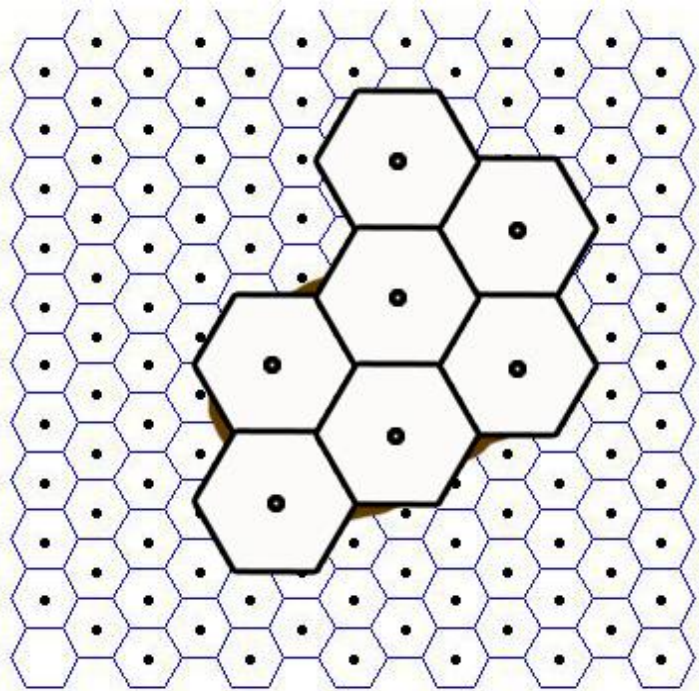

As the destiny of users on the mountain would be much lower than normal area, mountain area shares fewer carrier frequencies. A relative good solution requires only several very high antennae to divide the mountain area. Similar scheme used before can be applied here to handle overlap sections.

## Case 2: Mobile Vehicle Stations

For instance, a team of self-driving fans plan to travel past such a mountainous area with amateur VHF vehicle radio stations. Moreover, they just drive around the mountain instead of going onto the roof of mountain.

### Case 2.A: Small Mountains Straggling

Former repeaters that have not been covered by mountain area still keep unblocked communication among repeaters. So such a special case does not require any enhancement of hardware.

### Case 2.B: Large Mountain

In this case, it will very knotty for a driver to communicate with others if he is exactly in a hexagon without a repeater (covered by mountain). Naturally, we should build a repeater in the incomplete hexagon with its original carrier frequencies and PL code, as below:

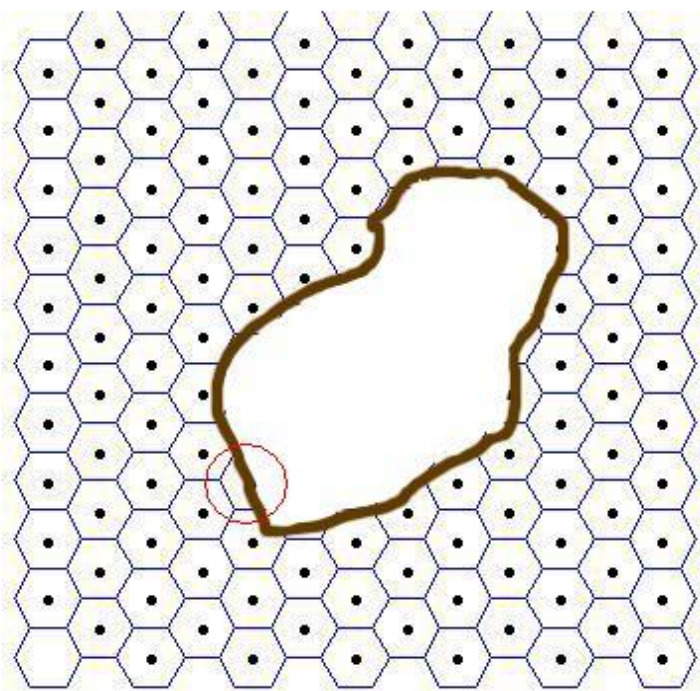

The hexagon in red circle is a unique area which needs a new repeater to be built.

# Model testing and sensitivity analysis

The following sheet shows the sensitivity of our model, every time we only change one variant and keep the others remaining. Cell mode and Group mode represent the 1000-user-model and the 10000-user-model we use above.

| variant | Cell mode | Group mode |
| --- | --- | --- |
| Population | Fits for small population size (less than 2187). Measures similar to the 1000-user-model can be adopted to analyze it. | Fits for the large popular size (more than 2187). Measures similar to the 10000-user-model can be adopted to analyze it. |
| The height of antenna H | H ↑, the square of a cell ↓, the number of total cells ↓, the number of repeaters ↓ | H has already been set to correspond with the area's communication request. |
| The frequency bandwidth $\Delta f$ | $\Delta f$ ↑, the number of carrier | $\Delta f$ ↑, the number of clusters |

| | frequencies ↓，the number of PL tones needed ↑ | ↑，a cluster's effective coverage ↓，the height of antenna H ↓，the number of groups ↓，the number of GC ↑. |
|---|---|---|
| The radius of the area supplied with radio service R | R ↑，the total number of cells ↑，the number of repeaters ↑.Notice: R cannot be too large; otherwise carrier frequencies are not enough. | The height of antenna H ↑. |

# Discuss the strength and weakness

## Strength:

- In our model, we use cellular structure to cover this area seamlessly with hexagonal cells or clusters. These cellular networks make it possible to cover with fewest repeaters.
- We use the repeaters that are capable of all the 54 PL tones. One only receives the cluster's specific PL tone but can retransmit any neighboring PL tones，while some older radios are only capable of a single PL tone.
- In the 1000-user-model, we divide this area into different clusters. And each cluster consists of some cells. We can make everyone in the 1000 users have a unique ID only by using the limited frequency channels and 54 PL tones.
- In the 10000-user-model, we give up the concept of cells and combine small clusters to make up groups. Every group has one unique group code (GC). With the help of GC, PL and specific frequency, people can find anyone they want to communicate.
- We also consider some practical situation in mountainous areas, which provide guidance to the practical application.

## Weakness and future work:

- In our model, no matter in the 1000-user-model or 10000-user-model, we assume that coverage of repeater is mainly decided by the height of antenna. But in the real application, the coverage of repeater is determined by all kinds of complex factors. Power of repeaters, location of repeaters, even the activity of sun would influence the coverage radius. Although

- these factors have already been noticed, more quantitative calculation may be introduced if we intend to build a more practical model.
- ➤ The communication between different cells and clusters is complex. Repeaters should memorize the information of how to get contact with anyone. If the number of users in this area is much more than 10000, there will be too much information to store. Trying to find an effective way to calculate the route without storing is one of the most important things in the future.
- ➤ Some information such as the group code and PL tone of objects is carried by the specific frequency of the objects. It is not convenient because the message the caller wants to transmit may be disturbed by this identification signal. We are going to find some measures to separate these signal.

# Conclusion

We make two cellular network models in order to make the area to be covered seamlessly. In the 1000-user-model, we separate this area into 91 cells which are determined by the coverage of the repeaters. Then we divide these 91 cells into 42 clusters, so each cluster includes one or three cells. For the sack of distinguishing every user in the area, one cluster can only contain 24 users, and these 24 users share one unique PL tone which is different from other clusters. When we come to the 10000-user-model, we give up the concept of "cells" because of the large quantity of the "clusters". In this case we combine numbers of clusters to make a "group". The unique ID for each group is the Group Code (GC). Each cluster in one group uses a unique PL tone. So we can ensure that repeaters can find anyone by both the GC and PL tones. We also provide the rules of communication between different cells or clusters. In the end we can make 1000 or 10000 users in this circular flat area of radius 40 miles radius use this communication system simultaneously.

We also list some practical application in the mountainous areas where there are defects in line-of-sight propagation. We discuss the different influence caused by mountains of different size, and then find suitable ways which are based on the models we make above to solve each problem.

# References

[1] http://www.metrocor.net/ctcss.htm    CTCSS (PL) Tones Frequencies Metropolitan Coordination Association, Inc.

[2] Continuous Tone-Coded Squelch System.    http://en.wikipedia.org/wiki/CTCSS

[3] Repeater.    http://en.wikipedia.org/wiki/Repeater

[4] Radio repeater.    http://en.wikipedia.org/wiki/Radio_repeater

[5] Very high frequency.    http://en.wikipedia.org/wiki/VHF

[6] Cellular network.    http://en.wikipedia.org/wiki/Cellular_network

[7] Line-of-sight propagation.    http://en.wikipedia.org/wiki/Line-of-sight_propagation

[8] CALCULATING REPEATER COVERAGE.    Information provided by Bill Smith Jr. KB6MCU.    http://www.tasma.org/coverage.pdf

[9] Ping Li, Relay Enhanced Cellular Networks: Frequency Planning and Protocol Design. Shanghai Jiao Tong University

[10] Theory of Wireless Communication Technology.    Network optimal center of Guangzhou branch office of China Mobile.